\documentclass[aps,prd,floatfix,superscriptaddress,onecolumn,preprintnumbers]{revtex4}  
\usepackage{amssymb}
\usepackage{amsmath}
\usepackage{amsfonts}
\usepackage{epsfig}             
\usepackage{graphicx}
\usepackage{stackrel}
\usepackage{tabularx}
\usepackage{color}
\newcommand{\seq}{\begin{subequations}}
\newcommand{\sen}{\end{subequations}}
\newcommand{\eq}{\begin{eqnarray}}
\newcommand{\en}{\end{eqnarray}}

\def\nn{\nonumber}

\begin{document}
\preprint{TTP21-049, RBI-ThPhys-2022-4}

\title{Lepton phenomenology of Stueckelberg portal to dark sector} 

\author{Aliaksei~Kachanovich} 
\affiliation{Institute for Theoretical Particle Physics (TTP), Karlsruhe Institute of Technology (KIT), 
Wolfgang-Gaede-Stra\ss e 1, 76131 Karlsruhe, Germany} 
\affiliation{Ruder Boskovic Institute, Bijenicka cesta 54, 10000 Zagreb, Croatia}
\author{Sergey~Kovalenko} 
\affiliation{Departamento de Ciencias F\'isicas, 
Universidad Andres Bello, Sazi\'e 2212, Piso 7, Santiago, Chile} 
\affiliation{Millennium Institute for Subatomic Physics at
the High-Energy Frontier (SAPHIR) of ANID, \\
Fern\'andez Concha 700, Santiago, Chile}
\affiliation{Centro Cient\'\i fico
	Tecnol\'ogico de Valpara\'\i so-CCTVal, \\ 
	Universidad T\'ecnica Federico Santa Mar\'\i a, Casilla 110-V, Valpara\'\i so, Chile}
\affiliation{Bogoliubov Laboratory of Theoretical Physics, JINR, 141980 Dubna, Russia} 
\author{Serguei~Kuleshov} 
\affiliation{Departamento de Ciencias F\'isicas, 
Universidad Andres Bello, Sazi\'e 2212, Piso 7, Santiago, Chile} 
\affiliation{Millennium Institute for Subatomic Physics at
the High-Energy Frontier (SAPHIR) of ANID, \\
Fern\'andez Concha 700, Santiago, Chile}
\author{Valery~E.~Lyubovitskij} 
\affiliation{Institut f\"ur Theoretische Physik, Universit\"at T\"ubingen, \\
		Kepler Center for Astro and Particle Physics, \\ 
		Auf der Morgenstelle 14, D-72076 T\"ubingen, Germany} 
\affiliation{Centro Cient\'\i fico
	Tecnol\'ogico de Valpara\'\i so-CCTVal, \\ 
	Universidad T\'ecnica Federico Santa Mar\'\i a, Casilla 110-V, Valpara\'\i so, Chile}
\affiliation{Millennium Institute for Subatomic Physics at
the High-Energy Frontier (SAPHIR) of ANID, \\
Fern\'andez Concha 700, Santiago, Chile}
\author{Alexey~S.~Zhevlakov} 
\affiliation{Bogoliubov Laboratory of Theoretical Physics, JINR, 141980 Dubna, Russia} 
\affiliation{Matrosov Institute for System Dynamics and 
Control Theory SB RAS, \\  Lermontov str., 134, 664033, Irkutsk, Russia } 

\date{\today}
		
\begin{abstract}

We propose an extension of the Standard Model (SM) with a $U_{A'}(1)$ gauge invariant dark dector 
connected to the SM via a new portal --- the Stueckelberg portal, arising in the framework of dark photon $A'$ mass generation via Stueckelberg mechanism.
This portal opens through the effective dim=5 operators constructed from the covariant 
term of the auxiliary Stueckelberg scalar field $\sigma$ providing flavor non-diagonal 
renormalizable couplings of both $\sigma$ and $A'$ to the SM fermions $\psi$. 
The Stueckelberg scalar plays a role of Goldstone boson in the generation of mass of the Dark Photon. 
Contrary to the conventional kinetic mixing portal,  in our scenario flavor diagonal  
$A'$-$\psi$ couplings are not proportional to the fermion charges and are, in general, flavor nondiagonal. 
These features drastically 
change the phenomenology of dark photon $A'$ relaxing or avoiding some previously established 
experimental constraints. We focus on the phenomenology of the described scenario of 
the Stueckelberg portal in the lepton sector and analyze the contribution of the dark sector 
fields $A'$ to the anomalous magnetic moment of muon $(g-2)_{\mu}$, 
lepton flavor violating decays $l_{i}\to l_{k}\gamma$ and $\mu-e$ conversion in nuclei. We obtain 
limits on the model parameters from the existing data on the corresponding observables.

\end{abstract}
	
\maketitle

\section{Introduction}

The idea of the dark sector (DS) of the Universe, existing almost independently of 
the Standard Model (SM) sector, has attracted growing interest in recent years. 
Originally, DS was thought to be populated by only one dark species, necessary to make up 
for the lack of matter in the Universe with dark matter (DM). 
In particular, extensions of DS were motivated by the popular scenario of  light sub-GeV DM.
It was realized that in this case a dark boson, known as the dark photon, would need 
to be introduced to prevent the Universe from overclosing. An extended DS can have not 
only cosmological, but also interesting phenomenological consequences. 
This DS physics beyond the SM can manifest itself in the phenomena observable experimentally 
(for a status report see, e.g., Ref.~\cite{Alexander:2016aln}).

Presently, there are a number of experiments to search for DS physics and others are planned 
for the near future. Among them, we mention CERN-based experiments 
NA64~\cite{Banerjee:2016tad,Banerjee:2017hhz}, 
NA62~\cite{CortinaGil:2019nuo}, 
SHiP~\cite{Alekhin:2015byh,Anelli:2015pba,SHiP:2020noy},  
LHCb~\cite{Aaij:2017rft}, ATLAS~\cite{Aad:2020arf}, and CMS~\cite{Sirunyan:2020zow} and {\it BABAR} experiment at SLAC~\cite{Lees:2017lec}, 
HPS at JLab~\cite{Adrian:2018scb}, and Belle at KEK~\cite{Park:2020lyz}. So far no signal of DS 
or another kind of new physics beyond the SM (BSM) is observed.  

An encouraging indication of new physics has recently come from measurements of the 
anomalous magnetic moment (AMM) of the muon $(g-2)_{\mu}$. The Fermilab Muon $g-2$ Collaboration 
published~\cite{Muong-2:2021ojo} the observation of 4.2 $\sigma$ deviation of the $(g-2)_{\mu}$ 
from its SM value and stimulated an explosion of the BSM literature.
As is known, measurements of $ (g-2)_{\mu} $ are a very sensitive probe of BSM physics. 
The Fermilab Muon $g-2$ result with such unprecedented precision can severely limit or refute 
many BSM models.

On the other hand, there is no doubt that the SM is an incomplete theory, requiring some physics 
beyond its scope to explain a number of problems that cannot be addressed in the SM. Among them, 
the DM problem is one of the most obvious. As we already mentioned, DM hints at the existence of 
a DS of the Universe, which not only provides DM particle candidates, but is also populated 
by other particles involved in interactions governed by some dark symmetries. The DS with possible 
nontrivial physics could have a phenomenological impact on the SM sector through portals such 
as the well-known kinetic mixing of dark and normal photons. Other hypothetical DS particles can 
have access to the SM sector through different portals and contribute to various observables, 
in particular, to $(g-2)_{\mu}$, allowing one to probe the DS.    

We should stress that there is much evidence of deviation from SM. 
Besides $(g-2)$ of muon~\cite{Aoyama:2020ynm,Dorokhov:2014iva},  
evidence includes the strong CP problem and rare meson decays~\cite{Pospelov:2017kep}-\cite{Zhevlakov:2020bvr},     
flavor nonuniversality~\cite{Ema:2016ops,Buras:2021btx},  $b-s$ quark anomaly, 
and others.  
It motivates theoretical study/construction of effective Lagrangians beyond Standard Model 
trying to involve new particles/portals, like the axion, dark photon, 
vectors, pseudoscalars, scalars, axials, etc.~\cite{Alexander:2016aln,Alekhin:2015byh},%
\cite{Kahlhoefer:2017dnp}-\cite{Buen-Abad:2021fwq}.  
 
Here, we propose an extension of the SM by inclusion of DS with $U_{A'}(1)$ symmetry. 
The corresponding gauge boson $A'$, also known as dark photon, acquires a nonzero gauge invariant 
mass via the Stueckelberg mechanism~\cite{Stueckelberg:1938zz,Ruegg:2003ps}, which implies the existence 
of a scalar Stueckelberg field $\sigma$ which is unphysical. 

This field opens a new portal from the SM to the dark sector
via the effective dimension-5 operator with the covariant derivative of the $\sigma$-field. 
We call it the Stueckelberg portal. In our setup this portal coexists with the conventional 
kinetic mixing portal and leads to new phenomenological effects in the SM sector, 
in particular, flavor violation both in lepton and quark sectors. 
In the present work we focus on the lepton flavor violation (LFV) and the corresponding experimental 
observables.

We also introduce one dark fermion, $\chi$, charged under $U(1)_{D}$, which is a viable light 
DM particle candidate. We postulate that dark scalar boson (DSB) plays important role in this model: 
(1) generating mass of the dark gauge boson (DGB) or dark photon, via the
Stueckelberg mechanism~\cite{Stueckelberg:1938zz,Ruegg:2003ps};  
(2) generate a mixing of DS with SM fermion including couplings preserving 
and violating symmetries of SM (e.g., lepton flavor violation (LFV)). 
Interaction of DGB and DSB with fermions is based on the idea of a familon 
(or flavons)~\cite{Feng:1997tn,Bauer:2019gfk,Cornella:2019uxs}. 
(3) plaing the role auxiliary field for DGB and reduce unphysical degrees of freedom of one.  

The paper is organized as follows. In Sec.~\ref{Sec2} we describe our theoretical setup.
In Secs.~\ref{Sec3},~\ref{Sec4}  and \ref{Sec5} we consider application of the Stueckelberg 
Portal to phenomenological aspects 
of the $g-2$ lepton anomaly, lepton conversion, and rare lepton-flavor-violating decays $l_i \to l_k \gamma$ 
which were used to derive limits for couplings occurring in the Stueckelberg portal.
In Sec.~\ref{Sec6} we discuss the boundary to  DGB  couplings and a possible contribution 
to $g-2$ lepton from obtained restrictions for different channels. Section~\ref{Conclusion} is the conclusion.
Technical details of our calculations are placed in Appendixes. 

\section{Theoretical setup}
\label{Sec2}
	
We consider an extension the SM with a $U_{A'}(1)$ gauge invariant dark sector described by the 
Lagrangian
\begin{eqnarray}\label{eq:Model-Lagr-1}
{\cal L}_{\rm SM+DS} = {\cal L}_{\rm SM} + 
{\cal L}_{\rm DS} + {\cal L}_{\rm port} \,, 
\end{eqnarray} 
where ${\cal L}_{\rm SM}$ and ${\cal L}_{\rm DS}$ are the SM and dark sector Lagrangians. 
Communication between these two sectors takes place through a portal interactions ${\cal L}_{\rm port}$.

We suppose that the DS, blind to the SM interactions, 
is populated with Dirac fermions $\chi_{i}$ charged under $U_{A'}(1)$. 
The lightest of which is stable and plays the role of dark matter. 
By definition, all the SM fields are neutral with respect to this group. 
The gauge boson, $A'$, of the dark sector $U_{A'}(1)$ group is called the dark photon.
In the conventional dark photon scenario $A'$ acquires its mass $m_{A'}$ from spontaneous breaking 
of the $U_{A'}(1)$ group.  
In contrast, in our approach its mass is a gauge-invariant quantity generated by 
the Stueckelberg mechanism. This requires the introduction of a scalar Stueckelberg field 
$\sigma$, which plays a role of Goldstone boson in the generation of mass of the dark photon. 
The $U_{A'}(1)$ gauge invariant Stueckelberg Lagrangian ${\cal L}_{\rm DS}$ 
with one specie of Dark fermion, $\chi$, reads 
\begin{eqnarray}\label{eq:DS-Lagr-1}
{\cal L}_{\rm DS} = - 
\frac{1}{4} \, {A}_{\mu\nu}' {A}^{\prime \mu\nu} 
\,+\, \frac{1}{2} D_\mu\sigma D^\mu\sigma 
\,+\, \bar\chi \, (i\not\!\!D_{\chi} - m_\chi) \, \chi  \,, 
\end{eqnarray}  
where,  as usual, $A_{\mu\nu}' = \partial_\mu A_{\nu}' - \partial_\nu A_{\mu}'$ and 
$\Big(iD_{\chi}\Big)_\mu = i\partial_\mu - g_{A'} A'_\mu$ 
is the conventional covariant derivative.  

The Stueckelberg covariant derivative is defined as
\begin{eqnarray}\label{eq:Covar-D-2}
D_\mu {\bf \sigma}    = \partial_\mu {\bf \sigma} - m_{A'} A'_\mu
\end{eqnarray} 
The $U_{A'}(1)$-symmetry is realized on the dark sector fields according to the transformations 
\begin{eqnarray}\label{eq:U1-transform-1}
A^\prime_\mu &\to& A^\prime_\mu + \frac{i}{g_{A'}} \, \partial_\mu\Omega_{A'} 
\, \Omega^{-1}_{A'} 
\,, \qquad A_{\mu\nu}' \,\to\, A_{\mu\nu}' \,, \\
\label{eq:U1-transform-2}
\sigma &\to& \sigma - \frac{m_{A'}}{g_{A'}} \theta_{A'}\,, \qquad 
\partial_\mu\sigma \,\to\, \partial_\mu\sigma 
+ \frac{i m_{A'}}{g_{A'}} \, \partial_\mu\Omega_{A'} \, \Omega^{-1}_{A'} 
\,, \qquad D_\mu\sigma \to D_\mu\sigma \,, \\
\label{eq:U1-transform-3}
\chi_D &\to& \Omega_{A'} \, \chi_D \,, \qquad 
i\not\!\! D_{\chi_D} \, \chi_D \,\to\, \Omega_{A'} 
\, i\not\!\! D_{\chi_D} \, \chi_D \,, 
\end{eqnarray} 
where 
\eq 
\Omega_{A'}(x) = \exp\Big[ i {\theta}_{A'}(x) \Big] \,.
\en 
As seen from (\ref{eq:U1-transform-2}), the $\sigma$ 
is an axionlike field shift transformed under the $U_{A'}(1)$.
 Quantization of  the $U(1)_{A'}$ dark sector requires adding to the Lagrangian  
(\ref{eq:DS-Lagr-1}) a gauge-fixing term \cite{Kors:2005uz}.  We choose it in the form
\begin{eqnarray}\label{eq:gf} 
\mathcal{L}_{gf}&=&  -\frac{1}{2\xi}\left(\partial_{\mu}A^{\prime \mu} +\xi m_{A'} \sigma\right)^{2},
\end{eqnarray}
where $\xi$ is a gauge parameter.
Then the dark sector Lagrangian can be written as
\begin{eqnarray}\label{eq:DS-Lagr-gf}
{\cal L}'_{\rm DS} = {\cal L}_{\rm DS} + \mathcal{L}_{gf}&=&  - 
\frac{1}{4} \, {A}_{\mu\nu}' {A}^{\prime \mu\nu} 
\,+\, \frac{m_{A'}^{2}}{2} A'_{\mu} A^{\prime \mu} 
\,+\, \bar\chi \, (i\not\!\!D_{\chi} - m_\chi) \, \chi   
-  \frac{1}{2\xi}\left(\partial_{\mu}A^{\prime \mu} \right)^{2} \nonumber\\
\nonumber
&+& \frac{1}{2} \partial_{\mu}\sigma\partial^{\mu}\sigma - \xi \frac{m_{A'}^{2}}{2} \sigma^{2} \,.
\end{eqnarray}  
In this gauge the Stueckelberg field is decoupled from other fields, making the theory manifestly unitary and renormalizable.  Note that the mass of the $\sigma$ field is proportional to the gauge parameter $\xi$, signaling that this field is unphysical.

In the gauge (\ref{eq:gf}) the dark photon propagator takes the form 
\begin{eqnarray}\label{eq:A'-propagator-1}
D_A^{\mu\nu}(k; \xi) = \frac{1}{k^2 - m_{A'}^2} \, 
\biggl[ g^{\mu\nu} - \frac{k^\mu k^\nu \, (1-\xi)}{k^2  - \xi m_{A'}^2} 
\biggr] \,.
\end{eqnarray}

Let us turn to the SM-DS portals ${\cal L}_{\rm port}$ possible in the present model.   
The well-known example of these is the generic renormalizable portal given by 
kinetic mixing of the dark and the SM photons, $A-A'$,  according to 
\eq\label{AAprime_mix}
{\cal L}_{\rm mix}  &=&
 - \frac{\epsilon_{A}}{2} \,   F_{\mu\nu} A^{\prime \mu\nu} \,, 
\en 
where $\epsilon_A$ is the mixing parameter. It has a rather particular phenomenology, 
which we comment on latter. 

In the Stueckelberg framework there exists another specific portal, which relies on  the 
$SU_{c}(3) \times SU_{L}(2) \times U_{Y}(1)\times U_{A'}(1)$ gauge-invariant effective operator
\begin{eqnarray}\label{eq:Stuckelberg-portal-1}
{\cal L}_{\rm int} = \frac{1}{\Lambda} \, 
D_\mu\sigma \,  \sum_{ij} \, 
\biggl[ 
\bar Q^{i} \chi^{ij} \gamma^\mu Q^{j}    \,+\, 
\bar u_R^i \chi^{ij}_{u} \gamma^\mu u_R^j        \,+\, 
\bar d_R^i \chi^{ij}_{d} \gamma^\mu d_R^j        \,+\, 
\bar L^{im} \kappa^{ij}_{L} \gamma^\mu L^{jm}     \,+\, 
\bar \ell_{R}^i \kappa^{ij}_{R} \gamma^\mu \ell_{R}^j 
\biggr]\,.
\end{eqnarray} 
The fields in this expression belong to the following representations of the 
$SU_{c}(3) \times SU_{L}(2) \times U_{Y}(1)\times U_{A'}(1)$ group 
$Q(3,2;1/3;0)$, $u_{R}(3,1;4/3;0)$, $d_{R}(3,1;-2/3;0)$, $L(1,2;-1;0)$, $\ell_{R}(1,1;-2;0)$. 
The parameter $\Lambda$ is the characteristic scale of this effective operator, defining  
when it opens up in terms of renormalizable interactions of an UV completion. 
The parameters $\chi^{ij}$ and $\kappa^{ij}$ form  
$3\otimes 3$ Hermitian matrices leading to the neutral current flavor violation both 
in quark and lepton sectors.  In the present work we focus only on the lepton sector and make an
ad hoc assumption $\chi=\chi_{u}=\chi_{d}=0$. 

Let us look closely at the effective operator~(\ref{eq:Stuckelberg-portal-1}). 
At first  glance, it looks a nonrenormalizable operator of dimension~5. However, after 
the substitution of the expression~(\ref{eq:Covar-D-2}), we find that the gauge-invariant  
operator~(\ref{eq:Stuckelberg-portal-1}) generates dimension-4 interactions of dark photon 
with the SM fermions $\psi$ in the form  
\begin{eqnarray}\label{eq:A'-psi-1} 
{\cal L}^{\rm A'-\psi}&=&  A'_{\mu}\sum_{ij}  \bar{\psi}_i \gamma^{\mu}\left(g^{V}_{ij} 
+ g^{A}_{ij}\gamma_5 \right) \psi_j \,,
\end{eqnarray}
where vector $g^{V}$ and axial-vector $g^{A}$ dimensionless couplings are defined as 
\begin{eqnarray}\label{eq:g-V-AV}
g_{ij}^{V}= \frac{m_{A'}} {\Lambda} v_{ij}  & \text{and} 
&  g_{ij}^{A}=\frac{m_{A'}} {\Lambda} a_{ij} \,, \\
v_{ij}=\frac{1}{2}\left(\kappa_{R}+ \kappa_{L}\right)_{ij}  
& \text{and} & a_{ij}=\frac{1}{2}\left(\kappa_{R}-  \kappa_{L}\right)_{ij} \nonumber 
\end{eqnarray}
As seen, these couplings are linearly scaled with the dark photon mass $m_{A'}$, 
which is crucial for our analysis of the $A'$ contribution to the lepton sector observables 
and setting limits on the corresponding couplings in function of the intermediate-state mass.

The operator~(\ref{eq:Stuckelberg-portal-1}) also contains the interaction 
of the unphysical Goldstone-like field with the SM fermions of the form $ j^{\mu}\partial_{\mu}\sigma$. 
In principle, these interactions should be taken into account in calculations made in 
the $R_{\xi}$ gauge~(\ref{eq:gf}) with the arbitrary parameter $\xi$. 
To avoid this complication we select from now on the unitary-type gauge 
$\xi\to\infty$ in which, as seen from~(\ref{eq:DS-Lagr-gf}), 
the $\sigma$ becomes infinitely heavy and decouples completely from the observable sector. 
Therefore, the only physical interactions generated by~(\ref{eq:Stuckelberg-portal-1}) 
in this gauge are due to the renormalizable couplings of the dark photon to 
the SM fermions~(\ref{eq:A'-psi-1}). 

These interactions also absorb the kinetic portal~(\ref{AAprime_mix}). 
In fact, the latter can be removed from the Lagrangian by the conventional field redefinition 
converting its effect to the flavor diagonal vector interactions of the form~(\ref{eq:A'-psi-1}). 
As usual, we shift the SM photon field
\eq\label{shift}
A_\mu \to A_\mu - \epsilon_{A} \, A_\mu^\prime  \, 
\en
and, then, rescale the dark photon field
\eq\label{rescaling}
A^\prime_\mu \to A^\prime_\mu  \, (1-\epsilon_{A}^2)^{-1/2}\,. 
\en 
These redefinitions generate flavor diagonal couplings of the SM fermions 
$\psi$ to the dark photon: 
\eq\label{LintSMD2}
{\cal L}_{\rm mix}^{\rm A'-\psi} =
e \, \epsilon_A \,  A^{\prime}_\mu   \, \bar\psi_{i} \, \gamma^\mu \, T_Q \, \psi_{i}  
\,,
\en
originating from the kinetic mixing term ~(\ref{AAprime_mix}).
Here $T_Q$ is the charge matrix of SM fermions.These interactions feature the characteristic property 
of the kinetic portal requiring the $A'$ couplings to the SM fermions to be proportional 
to their electric charges. 
It is clear that terms~(\ref{LintSMD2}) are completely absorbed by 
redefinition of the flavor diagonal vector couplings $g^{V}_{ii}$ in (\ref{eq:A'-psi-1}). 

Thus, our model contains the following free parameters $g_{ij}^{V}, \ g_{ij}^{A}, \  m_{A'}$ 
with $g_{ij}^{V,A} = g_{ji}^{V,A}$.  
Let us highlight two principal differences between the conventional kinetic portal and 
the Stueckelberg portal scenarios of dark sector. First, in the latter case contrary to 
the former one the $A'$ couplings to the SM fermions are not proportional to 
the SM fermion electric charges. Second, these couplings are flavor non-diagonal leading 
to reach LFV phenomenology.  
Note that the first point can significantly affect the conclusions following from the existing 
searches of dark photon.
In particular, the conventional 
dark photon from the kinetic portal scenario has been strongly constrained from the data of 
NA64 experiment at SPS CERN~\cite{Banerjee:2016tad,Banerjee:2017hhz}. For the Stueckelberg 
dark photon these constraints can be significantly relaxed. 

In the subsequent sections we will study contributions of the dark photon $A'$ to 
muon anomalous magnetic moment $(g-2)_{\mu}$ and LFV decays $l_{i} \to l_{k} \gamma$ as well as  
$\mu-e$ conversion in nuclei.

\section{Anomalous magnetic moment}
\label{Sec3}

\begin{figure}[t]
	\begin{tabular}{cc}
        \includegraphics[width=0.27\textwidth, trim={2cm 22.8cm 12cm 2cm},clip]{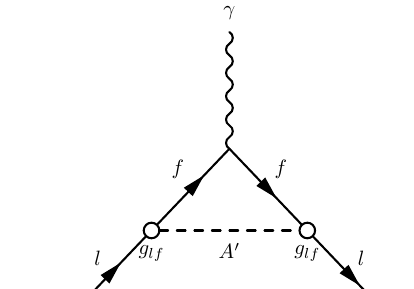}
	\end{tabular}
	\caption{Feynman diagrams describing the contributions of  
the Dark Sector vector $A'$ to the anomalous magnetic moments 
$\delta a_{l}$ of the leptons, taking into account flavor non-diagonal $l-f$ couplings, 
where  $l,f=e,\mu, \tau$.
}
	\label{fig:G2}
\end{figure}

In the Stueckelberg portal scenario the SM leptons $l$ receive the dark photon $A'$ 
one-loop contributions to their $(g-2)_{l}$ shown in Fig.~\ref{fig:G2}.  
The loop involves $A'$ due to its couplings to the  $l$ and $f$ SM fermions
according to Eqs.~(\ref{eq:A'-psi-1}). 
We calculate the corresponding the $A'$ loop contribution in the unitary gauge, setting 
$\xi\to\infty$ in~(\ref{eq:A'-propagator-1}). In this case the $\sigma$-field is decoupled from  
low-energy theory, as commented in the previous section. 

The first calculation of the vector and axial contributions 
to the lepton anomalous moments  in the $R_\xi$ gauge~(\ref{eq:A'-propagator-1}),  
taking into account LFV, was made in Ref.~\cite{Leveille:1977rc}. 
As it was noted in Ref.~\cite{Leveille:1977rc},  
the axial contributions $\delta a^{A}_{l}$ can obtained from the vector ones 
$\delta a^{V}_{l}$ by inverting the sign in front of the mass of the internal lepton 
$m_f \to - m_f$. In particular, the $\delta a^{V}_{l}$ and $\delta a^{A}_{l}$ 
contributions due to exchange of dark photon, $A'$, read~\cite{Leveille:1977rc}: 
\eq
\delta a^{V}_{l} &=&  \frac{(g_{lf}^V)^{2}}{4\pi^2} \, 
y_l \, 
\int\limits_0^1 \, dx \, \frac{1-x}{\Delta(x,y_A,y_l)} \,\Big[ 
x \ \Big(2 - y_l (1+x)\Big)  
\,+\, \frac{(1-y_l)^2}{2 y_A^2} \, (1 + y_l x) \, (1-x) \Big]\, 
\,, \label{aV_Rxi}\\ 
\delta a^{A}_{l} &=&  - \frac{(g_{lf}^{A})^{2}}{4\pi^2} \, 
y_l \, 
\int\limits_0^1 \, dx \, \frac{1-x}{\Delta(x,y_A,y_l)} \,\Big[ 
x \ \Big(2 + y_l (1+x)\Big)  
\,+\,\frac{(1+y_l)^2}{2 y_A^2} \, (1 - y_l x) \, (1-x) \Big] 
\,, \label{aA_Rxi} 
\en 
where we defined $y_l = m_l/m_f$, $y_A = m_{A'}/m_f$ 
and $\Delta(x,a,b) = a^2 x + (1-x) (1-b^2 x)$.    
The dimensionless couplings are defined in Eq.~(\ref{eq:g-V-AV}). 
For convenience we present details of the calculations of these integrals 
in Appendix~\ref{Rxi_aV_aA}.

\begin{figure*}[t!]
		\includegraphics[width=0.999\textwidth, trim={2cm 18.5cm 1cm 2.5cm},clip]{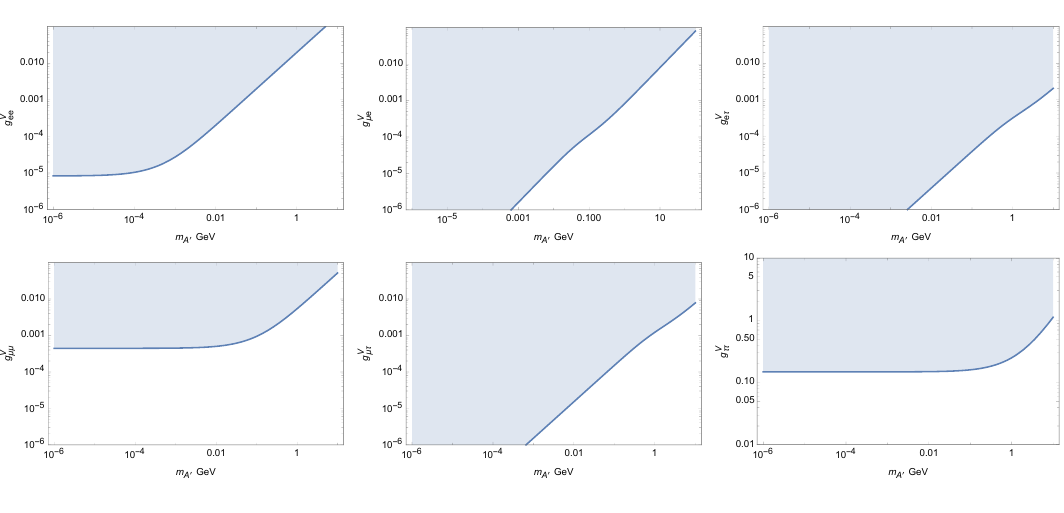}
	\caption{Upper bounds on the couplings $g^V$ of the dark photon $A'$  to the SM fermions 
                 as a function of its mass $m_A'$ derived from the data on leptonic $(g-2)$. 
		The shaded area is excluded.}
	\label{Bounds_LFV1}
\end{figure*}

\begin{figure*}[t!]
	\includegraphics[width=0.999\textwidth, trim={2cm 18.5cm 1cm 2.5cm},clip]{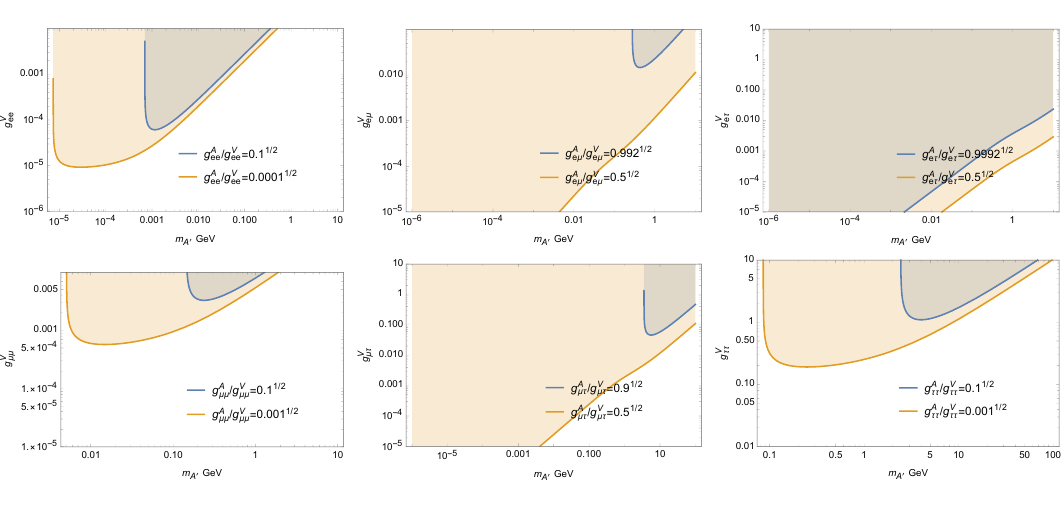}
	\caption{
	Upper bounds on the couplings $g^V_{ij}$ of the dark photon $A'$ to the SM fermions 
        in the vector+axial-vector channel as a function of its mass $m _{A'}$, 
        derived from the experimental data on $(g-2)_{l}$. The shaded area is excluded. 
        We show the plots for different values of the ratio $g_{ij}^{A}/g_{ij}^{V}$
		}\label{Bounds_LFV_AV}
\end{figure*}

The recent experimental measurements of the
anomalous magnetic dipole moments of  muon and electron $a_{\mu,e }=(g_{\mu,e}-2)/2$ 
demonstrate conspicuous deviation from the predictions of the SM. 
In particular, defining $\Delta a_{\ell} = a_{\ell}^{\mathrm{exp}}-a_{\ell}^{\mathrm{SM}}$,  
one gets 
\begin{eqnarray}
\Delta a_{\mu } &=&=\left(
2.51\pm 0.59\right) \times 10^{-9}~\mbox{\cite{Hagiwara:2011af,Davier:2017zfy,%
RBC:2018dos,Keshavarzi:2018mgv,Aoyama:2020ynm,Muong-2:2021ojo}}
\label{eq:a-mu} \\
\Delta a_{e} &=& (8.7\pm 0.5) \times   10^{-13}~\mbox{\cite{Keshavarzi:2019abf}}\,, \ \ \  
\Delta a_{e} \,=\, (4.8\pm 3.0)\times 10^{-13}~\mbox{\cite{Morel:2020dww}} \,. 
\label{eq:a-e}
\end{eqnarray} 
In the case of the muon, the value $a_{\mu }^{\mathrm{exp}}$ was extracted from the combined data of
the E821 experiment at BNL~\cite{Muong-2:2006rrc} and  recent Muon
$(g-2)$ measurements at FNAL~\cite{Muong-2:2021ojo}. This experimental result shows  4.2$\sigma$ deviation 
from the SM prediction. 
The value for $\Delta a_{e}$ was derived from the recent measurement of the fine-structure 
constant~\cite{Morel:2020dww}.  
For completeness we also include in our analysis the same observable for the $\tau$ lepton
\begin{eqnarray}\label{eq:a-tau}
\Delta a_\tau = (2.79 ) \times  10^{-4}~\mbox{\cite{Eidelman:2007sb,PDG20}}.
\end{eqnarray}
Its precision is significantly worse than for the case of $e$ and $\mu$. 
This is due to the experimental difficulties in measuring the properties of such a short-lived particle 
as the $\tau$ lepton. For rough estimations we will use the central value of $\Delta a_\tau$ in (\ref{eq:a-tau}).

We compare our theoretical predictions (\ref{aV_Rxi}) and (\ref{aA_Rxi}) with the experimental 
data~(\ref{eq:a-mu}), (\ref{eq:a-e}) and~(\ref{eq:a-tau}).  
First, we  extract upper limits on the coupling constants $g^{V,A}_{ij}$ of the dark photon 
to the SM fermions. Then, in Sec.~\ref{Sec6} we will discuss the possibility of simultaneous 
explanation of the muon and the electron $(g-2)$ in our model, taking into account the limits 
from LFV processes.

Extracting limits on $g^{V,A}_{ij}$ we apply the conventional simplifying assumption about 
the presence of only  one nonvanishing coupling constant at a time.  
In Figs.~\ref{Bounds_LFV1} and~\ref{Bounds_LFV_AV} we show the resulting upper limits 
for the couplings $g^{V,A}_{le}$, which are significantly more stringent than for other 
 combinations of flavor indices $g^{V,A}_{lf}$ with $f\neq e$  
due to the factor $m_{l}/m_{f}$ in Eqs.~(\ref{aV_Rxi}) and~(\ref{aA_Rxi}). 
These latter limits can be approximately obtained from $g^{V,A}_{le}$ with the corresponding 
rescaling using the mentioned factor.  

To estimate the effect of the combined contribution of $g^{V}$ and $g^{A}$ couplings we also 
studied the upper limits on the coupling $g^{V}$ for different values of the ratio $g^{A}/g^V$. 
The results are shown in Fig.~\ref{Bounds_LFV_AV}.

Let us to note that the combination of vector and negative axial-vector contributions  
to $(g-2)_\mu$ makes limits less stringent in comparison with the case of the pure vectorial term. 
This can significantly extend a window in the mass-coupling parameter space for the light dark vector particles.
Finally we note that, if the LFV effects in the $(g-2)$ were not taken into account, then 
the upper limits for the couplings $g_{ll}^{V}$ would be almost the same as for $g_{ll}^{A}$. 

\section{Nuclear Lepton Flavor Conversion}
\label{Sec4}
We recall again that the Stueckelberg portal model inherently features LFV couplings of the dark sector 
field $A'$ to the SM leptons, described by~(\ref{eq:A'-psi-1}).  
These LFV couplings can contribute to both flavor-conserving observables (e.g. $(g-2)_{l}$ 
studied in the previous section), and to the LFV ones, some of which we consider in what follows. 
We start with nuclear $\mu^{-}-e^{-}$ conversion, which is a LFV process with the participation of a nucleus, 
\begin{eqnarray}\label{eq:LFV-NC-1}
&&\ell_{1}^{-} + (A,Z)\to \ell_{2}^{-} + X.
\end{eqnarray}
It was recently advocated that deep inelastic lepton conversion on nuclei with $X$ denoting all 
the possible final-state particles, has good prospects for setting limits on the effective 
couplings of $\ell_{1}$ and $\ell_{2}$ 
in the $e-\mu,\tau$ and $\mu-\tau$ channels at the fixed-target 
NA64 experiment~\cite{Gninenko:2018num}. 

However, the process most studied experimentally is coherent $\mu^{-}-e^{-}$ conversion 
in muonic atoms, in which one electron is replaced by a muon. 
In this case $\ell_{1} = \mu$, $\ell_{2} = e$, and $X=(A,Z)$. 
This process has not yet been discovered experimentally. 
Presently the best upper limits on its rate $R_{\mu e}$ have been set by the SINDRUM II experiment 
on $\mu-e$ conversion in $^{198}$Au~\cite{Bertl:2006up}
\begin{eqnarray}\label{eq:SINDRUM-II}
&& R_{\mu e}^{{\rm Au}} \leq 4.3 \times 10^{-12}\, .
\end{eqnarray}
In the near future the PRISM/PRIME experiment~\cite{Witte:2012zza} with a titanium 
$^{48}$Ti target is going to reach the limit
\begin{eqnarray}\label{eq:PRISM}
&&R_{\mu e}^{{\rm Ti}}\  \lesssim 10^{-18}\, .
\end{eqnarray}
In Ref.~\cite{Gonzalez:2013rea} on the basis of nucleon-meson effective field theory, 
the above experimental 
limits were translated into the upper limits on the effective couplings $\alpha^{V,A}$ of 
the LFV $\mu-e$ current to nucleons defined by Lagrangian 
\begin{eqnarray}\label{eq:mu-e-N}
{\cal L}_{\rm eff}^{\ell N}  &=&\frac{1}{\Lambda_{LFV}^{2}} \bar N \gamma^\mu N 
\ \bar e \, \Big[\alpha^V \gamma_\mu + \alpha^A \gamma_\mu \gamma_5 \Big] \, \mu + 
{\rm H.c.}\,.
\end{eqnarray}
These limits are 
\eq 
&&\alpha_{A'}^{V,A} \left(\frac{1 \, \mbox{GeV}}{\Lambda_{\rm LFV}}\right)^2 \leq
8.5\times 10^{-13} \,,\hspace{5mm} \mbox{from SINDRUM II \ \ \ \cite{Bertl:2006up}},
\label{alpha-lim-Au}\\
&&\alpha_{A'}^{V,A} \left(\frac{1 \, \mbox{GeV}}{\Lambda_{\rm LFV}}\right)^2\leq  
1.6 \times 10^{-15} \, , \hspace{5mm}\mbox{from PRISM/PRIME~\cite{Witte:2012zza}.}
\label{alpha-lim-Ti}   
\en 
In our approach the Lagrangian~(\ref{eq:mu-e-N}) is generated at tree level by the $t$-channel exchange 
with $A'$ between the $\mu-e$ and $qq$ currents, where $q=u,d$  are valence quarks 
of the nucleon.  Note that transitions with $q_{1}\neq q_{2}$ do not contribute to coherent 
$\mu-e$ nuclear conversion. Starting from our Lagrangian~(\ref{eq:A'-psi-1}) 
and matching at a certain scale the quark currents with nucleon ones (see for details 
Ref.~\cite{Gonzalez:2013rea}) we find the relations 
\begin{eqnarray}\label{eq:alpha-g}
\alpha_{A'}^{V(A)} \simeq z\, g^{V(A)}_{qq} g^{V(A)}_{e\mu}\frac{m_{A'}^2}{m_{A'}^2+m_\mu^2} ,
	\end{eqnarray}
where $z$ is a dimensionless constant of order $\cal{O}$(1). 
Thus, from Eqs.~(\ref{alpha-lim-Au})-(\ref{eq:alpha-g}) we find 
the following upper limits:
\eq 
\label{eq:vv-lim-1}
\bigg|g^V_{e\mu}g^V_{qq}\bigg|\lesssim
\left\{ 
\begin{array}{cl}
8.5 \times 10^{-13}
\left(\dfrac{m_{A'}^2}{m_{A'}^2+m_\mu^2}\right)^{-1} & 
\ \ \ {\rm SINDRUM} \\[3mm]
1.6 \times 10^{-15}\, 
\left(\dfrac{m_{A'}^2}{m_{A'}^2+m_\mu^2}\right)^{-1}  & 
\ \ \ {\rm PRISM/PRIME} 
\end{array}
\right. 
\en 
We will use these limits in Sec.~\ref{Sec6} for our combined analysis of the $(g-2)_{l}$ 
and the LFV experimental data.

\section{LFV decays $l_i \to l_k \gamma$}
\label{Sec5}

\begin{figure}[b]		
	\includegraphics[width=0.79\textwidth, trim={2cm 22.5cm 1cm 2.5cm},clip]{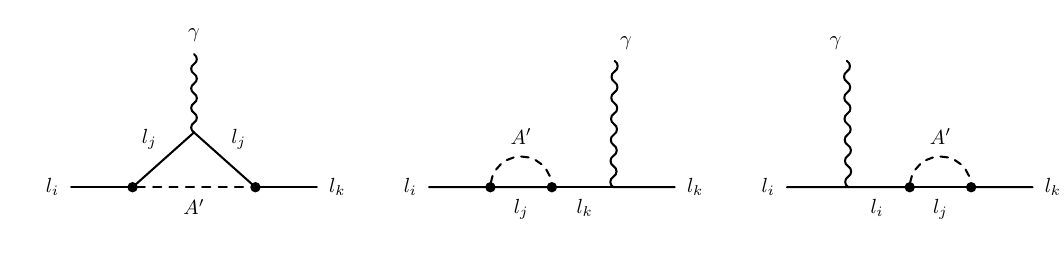}
	\caption{Feynman diagrams of gauge invariant matrix elements of the  interaction lepton with  
external electromagnetic field accounting LFV effect generated by the $A'$ dark photon.}
	\label{mue_LFV}
\end{figure}
The matrix element of this LFV process can be parametrized as
\eq
iM_{ik} = ie \, \epsilon^{\mu}(q) \, \bar{u}_k(p_2,m_k) \, \left[ \frac{i}{2m_i}\sigma_{\mu\nu}q^{\nu} 
F_M +\frac{i}{2 m_i}\sigma_{\mu\nu}q^{\nu}\gamma_5 F_D \right] \, u_i(p_1,m_i) \,. 
\en
Then
\eq
|M_{ik}|^2 &= & m_i^2 \, \biggl[1-\frac{m_k^2}{m_i^2}\biggr]^2 \, 
\biggl(|F_M|^2+ |F_D|^2\biggr) \,. 
\en
Here we used the Gordon identities listed in Appendix~\ref{App_A}.
Taking into account that $m_e \ll  m_\mu \ll m_\tau$, 
we have for the decay width of this process in a very good approximation the 
expression~\cite{Crivellin:2018qmi}
\eq
\Gamma(l_i \to l_k\gamma) &=& \frac{1}{2m_i} \int \frac{d^3p_2 \, d^3q}{4E_2 E_q  
\, (2\pi)^6} \, 
(2\pi)^4 \, \delta^{(4)}(p_1-p_2-q) \ |M_{ik}|^2  
\\
&=& \frac{\alpha}{2} \, m_i \, \biggl(|F_M|^2+ |F_D|^2 \biggr) \,, \nn
\en
where $F_M$ and $F_D$ are the magnetic 
and dipole form factors, and $\alpha =e^2/(4\pi) = 1/137.036$ is the fine-structure constant.  
The dark sector contributes to 
$l_i \to l_k \gamma$ decay only with the dark photon $A'$ according to the diagrams 
in Fig.~\ref{mue_LFV}. 

The corresponding analytical expressions for single LFV coupling read
\eq
F_M &=& - \frac{1}{16\pi^2} \left[\left(g^{V}_{ik}g^{V}_{ii}+ g^{A}_{ik}g^{A}_{ii}\right) 
h_2^{V}(x_\mu)+\left(g^{V}_{ik}g^{V}_{kk}+ g^{A}_{ik}g^{A}_{kk}\right)h_3^{V}(x_\mu)\right]\,, \\
F_D &=& \frac{1}{16\pi^2} \left[\left(g^{V}_{ik}g^{A}_{ii}+ g^{A}_{ik}g^{V}_{ii}\right) 
h_2^{V}(x_\mu)+\left(g^{V}_{ik}g^{A}_{kk}+ g^{A}_{ik}g^{V}_{kk}\right)h_3^{V}(x_\mu)\right]
\,. 
\en

For $\mu \to e \gamma$ process with the $\tau$ lepton in loop with double LFV coupling we have 
\eq
F_M &=&  -\frac{1}{16\pi^2} \left(\frac{m_\mu}{m_\tau}\right)\left[ g^{V}_{\mu \tau} 
g^{V}_{e \tau}h_1^{V}(x_\tau)+ g^{A}_{\mu \tau}g^{A}_{e \tau}h_1^{V}(x_\tau)\right]
\,, 
\nonumber\\
F_D &=& \frac{1}{16\pi^2} \left(\frac{m_\mu}{m_\tau}\right)\left[ g^{V}_{\mu \tau} 
g^{A}_{e \tau}h_1^{V}(x_\tau)+ g^{A}_{\mu \tau}g^{V}_{e \tau}h_1^{V}(x_\tau)\right] 
\,,
\en
where  $x_i = m_{A'}^2/m_i^2$. Expressions for the loop functions 
$h_i^{V}(x_i)$ in the approximation $m_e \ll m_\mu \ll m_\tau$ 
are shown in Appendix~\ref{App_D}. 

Let us note that the diagrams in Figs.~\ref{mue_LFV}(b) and~\ref{mue_LFV}(c) 
 are needed to guarantee gauge invariance of the photon interactions with leptons through 
loop diagrams induced by the $A'$  dark photon. This simultaneously leads to cancellation of 
a divergence arising from the diagram in Fig.~\ref{mue_LFV}(a). 

Similarly we can write the $A'$ contributions to $F_M$ and $F_D$ form factors of 
the for $\tau\to \mu \gamma$ 
and $\tau\to e \gamma$  LFV rare decays for the case when the initial or final leptons 
are different from the lepton in the loop. We have
for $\tau\to \mu \gamma$ decay
\eq
F_M &=& - \frac{1}{16\pi^2} \left[ g^{V}_{e\tau}g^{V}_{e\mu}+g^{A}_{e\tau}
g^{A}_{e\mu}\right]h_3^{V}(x_\tau) 
\,, \nonumber\\
F_D &=&  \frac{1}{16\pi^2} \left[ g^{V}_{e\tau}g^{A}_{e\mu}+ g^{A}_{e\tau}g^{V}_{e\mu}\right]
h_3^{V}(x_\tau)
\,, 
\en
and 
for $\tau\to e \gamma$ decay
\eq
F_M &=& 
- \frac{1}{16\pi^2} \left[ g^{V}_{\mu\tau}g^{V}_{e\mu}+ g^{A}_{\mu\tau}g^{A}_{e\mu}\right]h_3^{V}(x_\tau)
\,, \nonumber\\
F_D &=&  
\frac{1}{16\pi^2} \left[ g^{V}_{\mu\tau}g^{A}_{e\mu}+ g^{A}_{\mu\tau}g^{V}_{e\mu}\right]h_3^{V}(x_\tau)
\,.
\en 

\section{Analysis of current limits}
\label{Sec6}

In this section we derive experimental bounds on the $A'$ couplings $g^{V,A}_{ij}$ for several  
benchmark scenarios. 

\begin{figure}[t]		
	\includegraphics[width=0.99\textwidth, trim={2cm 21cm 1cm 2.7cm},clip]{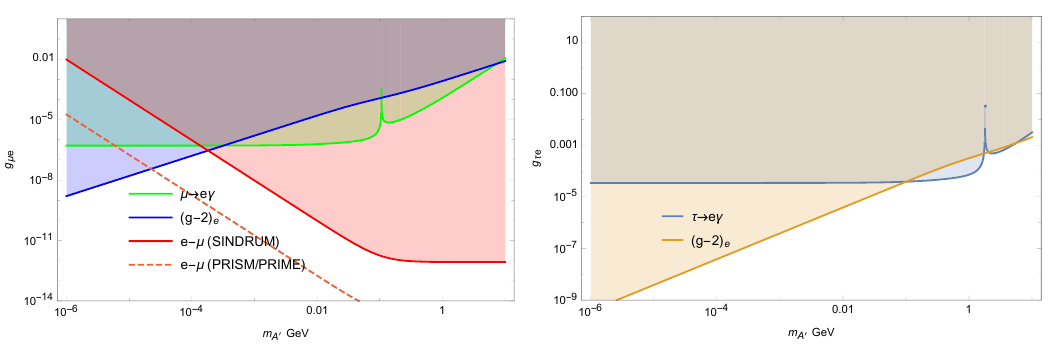}
	\caption{Limits on vector $g_{ij}$ couplings in dependence on masses 
$M_{A'}$ are deduced from an analysis of the following phenomena:  
$g-2$ ratios of leptons, widths of LFV decays $\mu\to e\gamma$ and $\tau\to e\gamma$ 
and lepton conversion. 
The shaded area is excluded by the data.}
	\label{Bound_LFV_and_g2}
\end{figure}

The current limits for the branchings of the LFV lepton decays $l_i \to l_k \gamma$ are~\cite{PDG20}  
\eq
\mathrm{Br}(\mu\to e\gamma ) &<& 4.2 \times 10^{-13}\,, \nonumber\\
\mathrm{Br}(\tau\to e\gamma ) &<& 3.3 \times 10^{-8}\,, \nonumber\\
\mathrm{Br}(\tau\to \mu\gamma ) &<& 4.4 \times 10^{-8} \,.
\en
First we analyze 
the dimensionless LFV couplings $g_{ij}$ by focusing on the scenario 
of lepton-flavor universality, assuming equal values of their diagonal elements, i.e.  
	$g^V_{ii}=g^{V}$ and $g^A_{ii}=g^{A}$ for $i=e,\mu,\tau$. 
	In this scenario we calculated and estimate coupling from $\mu\to e \gamma$ and $\tau\to e \gamma$ LFV decays.
	 The results are	presented in Fig.~\ref{Bound_LFV_and_g2} for the particular value $g_{ii}=1$.  
Bounds from the lepton $(g-2)_{l}$ (see  Fig.~\ref{Bound_LFV_and_g2}) are shown for the case of vanishing flavor-conserving couplings $g_{ii}$.   
Note that bounds from $\tau\to \mu \gamma$ are the same as for $\tau\to e \gamma$ because 
in the approximation 
$m_e \ll m_\mu \ll m_\tau$ the contributions from the loops are same. 
Also we omit in our analysis doubly LFV suppressed diagrams with heavy leptons propagating 
in the loop. 

The peaks in Fig.~\ref{Bound_LFV_and_g2} are induced by behavior of 
the loop integrals $h_i(x)$  near the point $x=1$ located in the vicinity of 
the vector boson production threshold. 
For resolving this problem one needs to include in our analysis 
finite width $\Gamma_{A'}$ of the decay of dark vector boson to the leptonic pair with 
$\Gamma_{A'} \sim \tau_{A'}^{-1} \sim g_{ij}^2$ in the Breit-Wigner propagator. 

Limits on the LFV couplings in Fig.~\ref{Bound_LFV_and_g2} include constraints from the lepton $(g-2)$ and 
rare LFV $l_i\to l_k\gamma$ lepton decays. In the case of $e-\mu$ LFV transition (see left pictures in 
Fig.~\ref{Bound_LFV_and_g2}) we add bound from $e-\mu$ conversion. 
Suppression for $e-\mu$ conversion at heavy masses is induced by heavy bosons exchange in the $t$ channel. 
The dark $A'$ photon keeps a window for possible huge LFV couplings $g_{ij}$ at light masses of leptons. 
We also would like to note that $g_{ij}$ is different from one will push the limits from LFV processes up.  

\begin{figure}[b]		
	\includegraphics[width=0.99\textwidth, trim={2cm 21cm 1cm 2.7cm},clip]{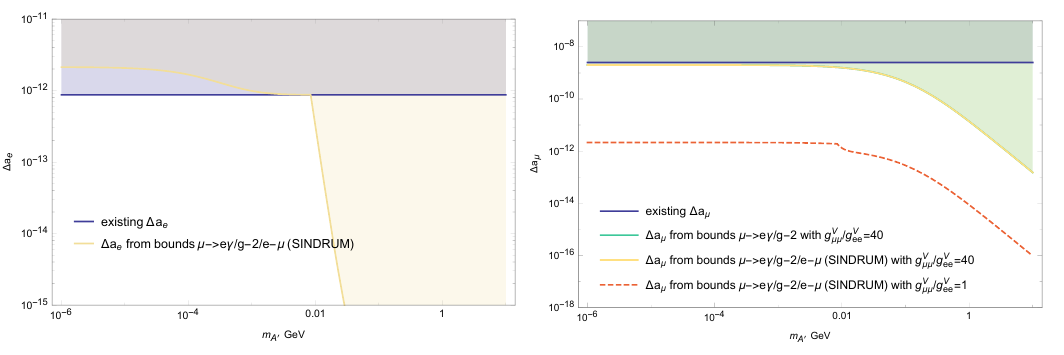}
	\caption{Estimate of contribution to lepton AMM is made in dependence on masses 
of dark photon. The limit is established for the benchmark case $g_{ee}=10^{-5}$ with taking into account the
restriction for LFV couplings}
	\label{g2EMu}
\end{figure}

Using the limit from  $l_i\to l_k \gamma$ decay in the scenario of universal lepton flavor-conserving (LFC) couplings $g^V_{ii}$ or $g^A_{ii}$ we can deduce the possible lepton contribution 
to $(g-2)$ in function of the mass of dark vector photon $A'$. 
We have estimated the contribution to $(g-2)_{l}$ of the sum of loops with light leptons
$e$ and $\mu$, taking into account the contribution of LFC and LFV  
couplings, which are constrained by $l_i\to l_k \gamma$ decay. In this way we write down 
\eq
\Delta a_{l_i} = (\Delta a_{l_i})_{\mathrm{LFC}}+ (\Delta a_{l_i})_{\mathrm{LFV}} \,.
\en
For the vector contribution we also included constraints from $\mu-e$ conversion. The results are shown in Fig.~\ref{g2EMu}. 
As can be seen, the dark photon $A'$ contribution to $(g-2)_{e}$ through the vector channel explains the electron anomaly $\Delta a_{e}$ for $m_{A'} < 10^{-2}$ GeV. 
Attempting to explain both $\Delta a_{e}$ and $\Delta a_{\mu}$ anomalies on account of  the $A'$ contribution, we find that it is not possible in at least in the lepton universal benchmark scenario where  $g_{\mu\mu}/g_{ee}=1$ (see the dashed line in the right panel in Fig.~\ref{g2EMu}). 
Going beyond this simplified scenario we can find the simultaneous solution of  $\Delta a_{e}$ and $\Delta a_{\mu}$ for $g^V_{\mu \mu} > g^V_{ee}$. A particular solution for   
$g_{\mu\mu}/g_{ee}=40$, properly taking into account the limits from the LFV processes (see Fig.~\ref{Bounds_LFV1}), is presented in the right panel of   Fig.~\ref{g2EMu}. We note, that this solution is pretty hierarchical requiring separation of two couplings of the similar nature in more than order of magnitude, which looks unnatural. 
Another possibility avoiding this kind of hierarchy would be to extent the field content of the model amending it with the dark sector (pseudo)scalars providing additional contributions to $(g-2)_{l}$. The study of this possibility is beyond the scope of this work. 

\section{Conclusions}
\label{Conclusion}

We constructed a phenomenological Lagrangian approach which combines SM and DM sectors 
based on the Stueckelberg mechanism for the generation mass of the dark  $U_D(1)$ gauge boson 
(or dark photon). The DM sector contains  dark photon, dark scalar, and generic dark
fermion fields. Note that the dark scalar generates the mass of the dark photon and plays a role of Goldstone boson in our gauge-invariant formalism.  Stueckelberg portal opens new possibilities for 
study of phenomenology of BSM physics and can be important for running and planning 
experiments at world-wide facilities (e.g., 
for the NA64 Experiment at SPS CERN~\cite{Banerjee:2016tad,Banerjee:2017hhz}).  

We derived the limits on the effective couplings of our Lagrangian using data 
on lepton AMMs, LFV lepton decays $l_i\to\gamma l_k$, 
and $\mu-e$ conversion. It is known that the latter are very useful because they give 
more stringent limits on the couplings of effective Lagrangian. 
We also found that the $(g-2)$ anomaly cannot be preferably solved within 
the Stueckelberg portal scenario by the light dark photon in the framework of conservative 
scenario with taking into account lepton universality. However, the simultaneous explanation of these both anomalies becomes possible once we allow approximately  one-order of magnitude hierarchy between the flavor diagonal couplings of $A'$ to electron $g_{ee}$ and to muon $g_{\mu\mu}$, which can be treated as moderately  unnatural. We mentioned the possible ways for relaxing this tension with the naturalness. We leave a detailed study of these aspects of our model for the future publications.

In the future we plan to study a possible role of the Stueckelberg portal in different LFV processes 
including semileptonic decays.  We plan include scalar, pseudoscalar dark bosons  
into the Stueckelberg portal of DM. We also intend to extended our ideas on non-Abelian scenario 
for the dark sector.
 
\begin{acknowledgments} 

This work was funded by BMBF (Germany) ``Verbundprojekt 05P2021 (ErUM-FSP T01) -
Run 3 von ALICE am LHC: Perturbative Berechnungen von Wirkungsquerschnitten
f\"ur ALICE" (F\"orderkennzeichen: 05P21VTCAA), by ANID PIA/APOYO AFB180002 (Chile),
by FONDECYT (Chile) under Grants No. 1191103 and No. 1190845 
and by ANID$-$Millennium Program$-$ICN2019\_044 (Chile).  
A.K.\ acknowledges support by DFG through
CRC TRR 257, \emph{Particle Physics Phenomenology after the Higgs Discovery} 
(Grant No.~396021762). The work of A.S.Z. is supported by a grant of AYSS JINR (Grant No. 22-302-02).

\end{acknowledgments}	

\appendix

\section{Contribution of Dark Photon to lepton anomalous magnetic moment in $R_\xi$ gauges}
\label{Rxi_aV_aA}

The most simple calculation of the contribution of dark photon to the 
lepton anomalous magnetic moment in $R_\xi$ gauges can be performed in the 
unitary gauge specified by the choice $\xi = \infty$. In particular, in this case 
contribution of Goldstone bosons explicitly vanishes. It is convenient 
to perform all calculations in dimensional regularization with $D=4-2\epsilon$ 
in order to explicitly show that 
all potentially divergent terms cancel each other leading to finite results. 
In particular, the sum of these terms is given by the integral: 
\eq 
I =\int\limits_0^1 dx \, \int \frac{d^Dk}{i \pi^{D/2}} \, \frac{k^2}{\Big[\Delta(x,y_A,y_l)-k^2\Big]^3} \, 
\frac{2}{D} \, \Big[4 - (D+2) (1-x)\Big] \,,
\en 
where $x$ is the Feynman parameter of integration. 

Performing integration of the loop integral in $D$-dimensions using master integral 
\eq 
I = \int\limits_0^1 dx \, \int \frac{d^Dk}{i \pi^{D/2} } \,
\frac{(k^2)^s}{\Big[\Delta-k^2\Big]^n} = \int\limits_0^1 dx \,
(-1)^s \, \frac{\Gamma(s+D/2) \, \Gamma(n-s-D/2)}{\Gamma(D/2) \, \Gamma(n) \,
(\Delta)^{n-s-D/2}}
\en
we get 
\eq 
I = - \int\limits_0^1 dx \, (1-x) \, \Big(4 - (D+2) (1-x)\Big) 
\, \Delta^{D/2-2}(y_A,y_l) \,.
\en 
Next using $D=4-2\epsilon$ and performing $\epsilon$ expansion 
\eq 
\Delta^{-\epsilon} = 1 - \epsilon \log[\Delta] 
\en 
we verify that the integral is finite and after straightforward simplifications 
at the limit $\epsilon \to 0$ is given by 
\eq
I = - \int\limits_0^1 dx \, \frac{(1-x)^2}{\Delta(x,y_A,y_l)} \, 
\Big[1 - y_l^2 x^2\Big] \,.  
\en 
Then summing all finite terms we arrive at the final results which are in full agreement with 
results of Ref.~\cite{Leveille:1977rc}.  For convenience, we display partial contributions 
to the integrand over Feynman parameter $x$, e.g., in case of dark photon connecting vector 
Dirac matrices $\gamma^\mu$ and $\gamma^\nu$. As was stressed in Ref.~\cite{Leveille:1977rc} 
and was pointed before in present manuscript the axial case is simply obtained from 
the vector case upon inverting a sign in front of the mass of external lepton$m_f$:

(1) contribution induced by transverse part of the dark photon propagator, i.e. 
by the $g^{\mu\nu}$ part: 
\eq 
\frac{x (1-x)}{\Delta(x,y_A,y_l)} \, \Big[2 - y_l (1+x)\Big]
\en 
(2) contribution induced by longitudinal part of dark photon propagator, i.e. 
by the $p^\mu p^\nu/m_{A'}^2$ part:  
\eq 
\frac{(1-x)^2}{\Delta(x,y_A,y_l)} \, \frac{(1-y_l)^2}{2 y_A^2} \, (1 + y_l x) \,. 
\en 

\section{Gordon Identities}
\label{App_A}
The Gordon identities for the matrix elements describing the coupling of  the
external gauge field with fermions having different masses read: 

\eq
i\sigma_{\mu\nu}q^\nu &=& -P_\mu +(m_i+m_j)\gamma_\mu \,, 
\qquad\qquad\quad
i\sigma_{\mu\nu}P^\nu = -q_\mu+(m_j-m_i)\gamma_\mu \,, \\
i\sigma_{\mu\nu}q^\nu\gamma_5 &=& -P_\mu\gamma_5 +(m_j-m_i)\gamma_\mu\gamma_5 \,, 
\qquad
i\sigma_{\mu\nu}P^\nu\gamma_5 =-q_\mu\gamma_5+(m_i+m_j)\gamma_\mu\gamma_5 \,, 
\en
where $P=p_1+p_2$ and $q=p_1-p_2$

\section{Diagrams in Figs.~\ref{mue_LFV}}
\label{Fig6cde}

In this Appendix we explicitly demonstrate that the set of three diagrams in
Figs.~\ref{mue_LFV}(a)-(c) is gauge invariant and finite.
To simplify the proof of gauge invariance we split the contribution of each diagram $M$ from
the set~\ref{mue_LFV} into a part which is manifestly gauge invariant $M^{\rm GI}$
and and one which is not $M^{\rm rest}$ as $M = M^{\rm GI} + M^{\rm rest}$.
This separation can be achieved in the following manner.
For the $\gamma$-matrix contacting with photon field we use the
substitution $\gamma^\mu = \gamma^\mu_\perp + q^\mu\!\not\!\! q/q^2$
where $\gamma^\mu_\perp = \gamma^\mu - q^\mu\!\not\!\! q/q^2$
obeys the transversity condition $\gamma^\mu_\perp \cdot q_\mu = 0$.
Expressions for diagrams containing only $\perp$-values
are gauge invariant separately. It is easy to show that the remaining terms, which are not gauge
invariant, cancel each other in total (see the detailed discussion of this technique, e.g.,
in Ref.~\cite{Faessler:2003yf}).
Therefore, it is enough to consider only the sum of the gauge-invariant contributions from
all diagrams in Fig.~\ref{mue_LFV}(a)-4(c). The proof of cancellation of the remaining terms
is straightforward. Below we list these terms for each diagram using dimensional regularization 
and for general case of exchange by the boson particle (with spin 0 or 1), i.e. we do not restrict 
to the exchange of dark photon: 
\eq
M^{\rm rest}_{4a} &=& \int \frac{d^Dk}{(2\pi)^D i} \,
\Gamma_1 \, \frac{1}{\not\! p' + \not\! k - m_f} \, \frac{q^\mu \not\! q}{q^2} \,
\frac{1}{\not\! p + \not\! k - m_j}  \, \Gamma_2 
\, \frac{d^{\Gamma_1\Gamma_2}(k)}{k^2 - m^2} \,,
\label{fig6c_II}\\
M^{\rm rest}_{4b} &=& \frac{q^\mu \not\! q}{q^2} \,
\frac{1}{\not\! p - m_k} \,
\int \frac{d^Dk}{(2\pi)^D i} \,
\Gamma_1 \, \frac{1}{\not\! p + \not\! k - m_j} \, \Gamma_2 
\, \frac{d^{\Gamma_1\Gamma_2}(k)}{k^2 - m^2} \,,
\label{fig6d_II}\\
M^{\rm rest}_{4c} &=&
\int \frac{d^Dk}{(2\pi)^D i} \,
\Gamma_1 \, \frac{1}{\not\! p' + \not\! k - m_j} \, \Gamma_2 
\, \frac{d^{\Gamma_1\Gamma_2}(k)}{k^2 - m^2}  \,
\frac{1}{\not\! p' - m_i} \, \frac{q^\mu \not\! q}{q^2} \,.
\label{fig6e_II}
\en 
Here $\Gamma_1$ and $\Gamma_2$ are the corresponding Dirac matrices;  
$d^{\Gamma_1\Gamma_2}(k) = 1$ for exchange by 
scalar/pseudoscalar particles with 	
$\Gamma_1 = I, \gamma^5$ and $\Gamma_2 = I, \gamma^5$ 
and $d^{\mu\nu}(k) = - g^{\mu\nu} + k^\mu k^\nu/m^2$ for exchange 
by vector/axial particles with 
$\Gamma_1 = \gamma^\mu, \gamma^\mu\gamma^5$  
and $\Gamma_2 = \gamma^\nu, \gamma^\nu\gamma^5$.  

Next, using the Ward identity for inverse fermion propagators 
$\not\! q = (\not\! p - m_\ell) - (\not\! p' - m_\ell)$ 
and free Dirac equations of motion for initial and final leptons
we simplify expressions for the individual rest matrix elements as:
\eq
M^{\rm rest}_{4a} &=& \int \frac{d^Dk}{(2\pi)^D i} \, \Gamma_1 \, 
\biggl[ \frac{1}{\not\! p' + \not\! k - m_j}
- \frac{1}{\not\! p + \not\! k - m_j}
\biggr] \, \Gamma_2 \, \frac{d^{\Gamma_1\Gamma_2}(k)}{k^2 - m^2}\,,
\label{fig6c_III}\\
M^{\rm rest}_{4b} &=&\int \frac{d^Dk}{(2\pi)^D i} \, \Gamma_1
\, \frac{1}{\not\! p + \not\! k - m_j}
\, \Gamma_2 \, \frac{d^{\Gamma_1\Gamma_2}(k)}{k^2 - m^2} \,,
\label{fig6d_III}\\
M^{\rm rest}_{4c} &=& - \int \frac{d^Dk}{(2\pi)^D i} \, \Gamma_1
\, \frac{1}{\not\! p' + \not\! k - m_j}
\, \Gamma_2 \, \frac{d^{\Gamma_1\Gamma_2}(k)}{k^2 - m^2} \,. \label{fig6e_III} 
\en
Finally, summing Eqs.~(\ref{fig6c_III})-(\ref{fig6e_III}) we get 0, therefore,
proving gauge invariance of the set~\ref{mue_LFV}(a)-\ref{mue_LFV}(c).

Now we turn to the discussion of finiteness of the sum of the set of diagrams~\ref{mue_LFV}(a)-(c).
The logarithmically divergent term in diagram~\ref{mue_LFV}(a) is generated by the part
of numerator containing two loop momenta: $\!\not\! k \, \gamma^\mu \!\not\! k$.
Applying dimension regularization with $D=4-2\epsilon$ it gives the following divergent result 
in case of exchange by a scalar $S$ or pseudoscalar $P$ particle with $\Gamma_1 = \Gamma_2 = I$ 
or $i \gamma^5$: 
\eq
M^{UV; S/P}_{4a} = - \frac{\gamma^\mu}{2 \epsilon} \,.
\en
The diagrams in Figs.~\ref{mue_LFV}(d) and~\ref{mue_LFV}(e) induce the following logarithmic divergencies:
\eq
M^{UV; S/P}_{4b} = \frac{\gamma^\mu}{2\epsilon}  \, \frac{m_i \pm  2 m_j}{m_i- m_k}\,.
\en
and
\eq
M^{UV; S/P}_{4c} = \frac{\gamma^\mu}{2\epsilon}  \, \frac{m_k \pm 2 m_j}{m_k- m_i}\,.
\en
respectively. Here and below $\pm$ corresponds to exchange of a 
scalar or pseudoscalar particle.. 

Summing up divergent contributions of three diagrams we get exact cancellation of the latter:
\eq
M^{UV; S/P}_{4a} + M^{UV; S/P}_{4b} + M^{UV; S/P}_{4c} = \frac{\gamma^\mu}{2\epsilon} \,
\biggl[ -1 + \frac{m_i \pm 2 m_j}{m_i- m_k}
+ \frac{m_k \pm 2 m_j}{m_k- m_i} \biggr] = 0 \,. 
\en
In case of the diagrams induced by vector particle exchange the logarithmic divergences 
induced by individual diagrams read (we explicitly show the contribution of 
transverse and longitudinal part of the vector $V$ or axial $A$ propagator, 
which are supplied by the subscript $T$ and $L$, respectively) 
\eq
M^{UV; V/A}_{4a} &=& M^{UV; V/A}_{4a; T} + M^{UV; V/A}_{4a; L} 
\,, \nonumber\\
M^{UV; V/A}_{4a; T} &=& - \frac{\gamma^\mu}{\epsilon} \,, \\
M^{UV; V/A}_{4a; L} &=& \frac{\gamma^\mu}{\epsilon} \, 
\biggl[ 1 + \frac{3}{2} \, \frac{m_j^2}{m^2} 
- \frac{m_i^2+m_k^2+m_i m_k}{2 m^2} 
\biggr]\,, \nonumber
\en 
\eq 
M^{UV; V/A}_{4b} &=& M^{UV; V/A}_{4b; T} + M^{UV; V/A}_{4b; L} \,, 
\nonumber\\
M^{UV; V/A}_{4b; T} &=& \frac{\gamma^\mu}{\epsilon} 
\, \frac{m_i \mp 2 m_j}{m_i-m_k}
\,, \\
M^{UV; V/A}_{4b; L} &=& \frac{\gamma^\mu}{\epsilon} 
\, \frac{m_k \mp m_j}{m_i-m_k}
\biggl[ 1 + \frac{m_j^2}{m^2} 
- \frac{m_k (m_k \pm m_j)}{2 m^2} 
\biggr]\,,  \nn
\en 
\eq 
M^{UV; V/A}_{4c} &=& M^{UV; V/A}_{4c; T} + M^{UV; V/A}_{4c; L} \,, 
\nonumber\\
M^{UV; V/A}_{4c; T} &=& \frac{\gamma^\mu}{\epsilon} 
\, \frac{m_k \mp 2 m_j}{m_k-m_i}
\,, \\
M^{UV; V/A}_{4c; L} &=& \frac{\gamma^\mu}{\epsilon} 
\, \frac{m_i \mp m_j}{m_k-m_i}
\biggl[ 1 + \frac{m_j^2}{m^2} 
- \frac{m_i (m_i \pm m_j)}{2 m^2} 
\biggr] \,. \nonumber
\en
Here $\pm$ corresponds to exchange of a 
vector or axial particle.

It is easy to show that in case of vector and particle exchange  we also get exact 
cancellation of the divergences and it occurs separately for transverse and 
longitudinal contribution of the propagator of the exchange particle: 
\eq 
& &M^{UV; V/A}_{4a} + M^{UV; V/A}_{4b} + M^{UV; V/A}_{4c} = 0 \,, 
\nonumber\\
& &M^{UV; V/A}_{4a; T} + M^{UV; V/A}_{4b; T} + M^{UV; V/A}_{4c; T} = 0 \,, \\
& &M^{UV; V/A}_{4a; L} + M^{UV; V/A}_{4b; L} + M^{UV; V/A}_{4c; L} = 0 \,. 
\nonumber
\en 

\section{Loop functions $h^{V}_1(x)$}
\label{App_D}

In this Appendix, we present the analytical expressions of the loop integrals occurring in 
the amplitude of the LFV decays $l_i \to\gamma l_k$ for different channels 
and leptons propagating in the loop in the approximation $m_e \ll m_\mu \ll m_\tau$: 
\eq
h^{V}_1(x) 
&=& -\frac{(4x^3-3x^2-6x^2\ln(x)-1)}{x(1-x)^3} \,, 
\\
h^V_2(x) &=&   
2 \left(2 \text{Li}_2(1-x)-2 \text{Li}_2\left(\frac{2}{-x+\sqrt{(x-4) x}+2}\right)+2
\text{Li}_2\left(\frac{2}{x+\sqrt{(x-4) x}}\right)-2 x \right.
\\&+& \left. \log
^2\left(\frac{x+\sqrt{(x-4) x}}{2 x}\right)+\frac{(x+1) ((x-4) x+2) \log
	\left(x\right)}{x-1}-2 x \sqrt{(x-4) x} 
\log \left(\frac{\sqrt{x}+\sqrt{(x-4) }}{2}\right)+1\right) \,, \nn
\\
h^{V}_3(x) &=&-4 x+4 (x-1)^2  \ln \left(\frac{x}{x-1}\right)+6 \,.
\en
All results have been numerically and analytically 
cross-checked using the Mathematica Package-X~\cite{Patel:2015tea} and packages  
FeynHelpers~\cite{Shtabovenko:2016whf} and FeynCalc~\cite{FeynCalc}.

\end{document}